\documentclass{elsart}
\usepackage{graphicx,amssymb}

\usepackage{epsfig}

\newcommand{\ket}[1]{\ensuremath{|#1 \rangle}}

\begin{document}

\begin{frontmatter}
\title{Disentanglement in qubit-qutrit systems.}

\author[TUD]{Mazhar Ali\corauthref{cor}},
\corauth[cor]{Corresponding author}
\ead{mazhar.ali@physik.tu-darmstadt.de}
\author[USA]{A R P Rau} and 
\author[TUD]{Kedar Ranade}

\address[TUD]{Institut f\"{u}r Angewandte Physik, Technische Universit\"{a}t Darmstadt, D-64289 Darmstadt, Germany}
\address[USA]{Department of Physics and Astronomy, Louisiana State University, Baton Rouge, Louisiana 70803, USA}

\begin{abstract}
We examine the phenomenon of {\it entanglement sudden death} (ESD) for $(2 \times 3)$-dimensional systems. As for $2 \times 2$ systems,  the negativity vanishes in finite time for some entangled pure as well as mixed states. While locally equivalent pure states do so asymptotically. Interference between the decay of the two upper levels to the lowest one in the qutrit adds further richness to ESD in this systems.
\end{abstract}


\end{frontmatter}

\section{Introduction}

Quantum entanglement has been recognized as the vital resource for the applications of \emph{quantum information and quantum computation} \cite{NC00}. Most of these applications require maximally entangled pure states such as the Bell states. However, due to the interaction with the environment, we have more often mixed states instead of pure states. This interaction between the closed quantum systems and the environment degrades the amount of entanglement initially present. Hence it is important to understand the dynamics of entanglement during the interaction of closed quantum systems with the environment. It is also important to look for those mixed entangled states which are robust against decoherence effects and are still useful for quantum technology. The last two decades have witnessed a tremendous amount of work in this fascinating and counter-intuitive phenomenon of entanglement. Many aspects of entanglement have been understood and many useful criteria worked out to detect and quantify the amount of entanglement present in physical bipartite as well as multipartite states. For more details, see the recent review by Horodecki {\it et al.} \cite{HHH07}.

It is well known that the interaction of the principal system (system of concern) with the environment (any other system whose dynamics we want to neglect) leads to decoherence. This decoherence is of concern in quantum information processing in that it gradually degrades the quantum correlations present. Indeed, it is not surprising that the decoherence takes infinite time to erase completely the quantum correlations in a physical system. The striking issue is that, although it takes infinite time to complete the decoherence locally, nevertheless the global entanglement can vanish in some finite time. Such a phenomenon has been named {\it entanglement sudden death} (ESD). Initially, Yu and Eberly reported this effect for two-qubit entangled states \cite{YE04}. After that many groups have reported this effect in different contexts and in higher dimensions of Hilbert space \cite{YE06,J06,DJ06,LR07,SW07,EY07,AM07,AQJ07}. In all previous work, except for a recent one \cite{AJ07} nearly simultaneous with ours, the subsystems of the principal physical system are of equal dimensions in Hilbert space. So it is important to investigate this effect also in subsystems of unequal dimensions, as, for example, a qubit-qutrit system. Our analysis indicates that ESD can happen also in them and in all dimensions of Hilbert space. 

This work is organized as follows: In section 2, we study the measures of entanglement that are suitable for our problem. In section 3, we model our system in the domain of quantum optics and study the dynamics of entangled pure states in the presence of decoherence. The two- and three-level atoms serve as qubit and qutrit, respectively. The atoms are the system of interest and the traps containing them are considered as the environment. In section 4, we investigate the behavior of entanglement in particular mixed states and report the existence of ESD in them. Finally, we conclude our work and indicate future directions in section 5.

\section{Measures of entanglement and maximally entangled pure states for the $2\times3$ systems}

The first step is to choose the measures of entanglement for entangled states of the $2 \times 3$ system. Unfortunately, except for the two-qubit case, there does not exist any operational method to compute all known entanglement measures for general mixed states of higher dimensions of Hilbert space. Negativity \cite{W02} is a measure, which is computable in all dimensions of Hilbert spaces for all the entangled states having a negative partial transpose (NPT). It is well known that for $2\times2$ and $2 \times 3$ Hilbert spaces, all states having positive partial transpose (PPT) (therefore, zero negativity) are separable \cite{HH96, P96}. Therefore, for our qubit-qutrit system of interest, negativity serves the purpose.

We can define the negativity as: {\it twice the absolute sum of all the negative eigenvalues of the partial transpose of a quantum state}. This measure is easy to compute and its value varies from zero for PPT states (hence separable for our system) to one for the maximally entangled states. Therefore, negativity is a reasonable entanglement measure for the qubit-qutrit entangled states.

The second step is to look for the maximally entangled pure states in the $2\times3$ Hilbert space. Such states are obviously extensions of the Bell states. In order to describe the set of pure entangled states in the six-dimensional Hilbert space, we can define an arbitrary pure state as

\begin{eqnarray} \label{eq:1}
\ket{\Psi} = \sum_{ij} \, a_{ij}\, \ket{i,j},
\end{eqnarray}
where $i$ takes the values $ 0$ and $1$ to denote the qubit states, and $j$ the values $0, 1,$ and $2$ to represent the qutrit state. The normalization of the state vector demands that $\sum_{ij} |a_{ij}|^2 = 1$. The $a_{ij}$s are complex numbers. The density operator $\rho$ for this pure state is defined as $\rho = |\Psi\rangle\langle\Psi|$. Now we take the partial transpose of the matrix $\rho$ with respect to the qutrit basis (taking partial transpose with respect to the qubit basis gives the same result). The spectrum of the partially transposed matrix $\rho^{T_{B}}$ is given by 
\begin{eqnarray} \label{eq:2}
\{0, \, 0, - \sqrt{f(a_{x})}, \, \sqrt{f(a_{x})}, \, \frac{1}{2}(1-\sqrt{1- 4 f(a_{x})}), \, \frac{1}{2}(1 + \sqrt{1 - 4 f(a_{x})} \}, 
\end{eqnarray}
where $a_{x} = a_{ij}$ and $ x $ takes values $ 1 $ to $ 6 $ corresponding to the six combinations of $a_{ij}$. i.e. $a_{1} = a_{00}$, $a_{2} = a_{01}$, $\ldots$, $a_{6} = a_{12}$.

Surprisingly, all the eigenvalues of the partially transposed matrix depend on the single function $f(a_{x})$ of the parameters $a_{x}$. The expression for $f(a_{x})$ is given by 

\begin{eqnarray} \label{eq:3}
f(a_{x}) &=& |a_{2}|^2 |a_{4}|^2 + |a_{3}|^2 |a_{4}|^2 - 2 \mathrm{Re} (a_{1} \bar{a}_{2} \bar{a}_{4} a_{5}) - 2 \mathrm{Re}(a_{1} \bar{a}_{3} \bar{a}_{4} a_{6}) \nonumber \\ &&+ |a_{1}|^2 \, |a_{5}|^2  + |a_{3}|^2 |a_{5}|^2 + |a_{1}|^2 |a_{6}|^2 + |a_{2}|^2 |a_{6}|^2 - 2 \mathrm{Re}(a_{2} \bar{a}_{3} \bar{a}_{5} a_{6}).
\end{eqnarray}

As the partially transposed matrix of a Hermitian operator is also Hermitian, all the eigenvalues must be real. This implies that $ 0 \leq f(a_{x}) \leq \frac{1}{4}$. At one extreme of the lower bound, the pure states are separable, while for the other extreme of upper bound, the pure states are maximally entangled. For $ 0 < f(a_{x}) < \frac{1}{4}$, the pure states are non-maximally entangled states. For the pure state to be entangled, at least two coefficients in eq(\ref{eq:1}) must be non-zero.

It is evident from the spectrum of the partial transpose of this class of pure states that all pure entangled states for the qubit-qutrit system can have only one negative eigenvalue. Therefore the negativity for such pure entangled states can be written as 

\begin{eqnarray} \label{eq:4}
N = 2 \sqrt{f(a_{x})}. 
\end{eqnarray}
As the negativity is invariant under all local unitaries, the square root of the function $f(a_{x})$ is also invariant under all local unitaries. 

\par By the Schmidt decomposition, an arbitrary qubit-qutrit state may be written in the form

\begin{equation}\label{eq:5}
\ket{\Phi_{1}} 
                 = (U_A \otimes U_B)\left(\alpha \ket{0, 0} + \sqrt{1 - \alpha^2} \ket{1, 1}\right),
\end{equation}
\noindent where $U_A$ and $U_B$ denote the transformation from the computational basis to the Schmidt basis on the
qubit and the qutrit, respectively, and where $\alpha \in [0;1/\sqrt{2}]$. where $ \ket{0, 0} = \ket{0}_{A} \otimes \ket{0}_{B}$, etc. The first vectors $\ket{0}$ and $\ket{1}$ are the orthonormal vectors for Alice's (qubit) Hilbert space, while the second vectors $\ket{0}$, $\ket{1}$, and $\ket{2}$ are for Bob's (qutrit) Hilbert space. Since $f$ is invariant with respect
to local unitary operations, we may ignore $U_A$ and $U_B$ for now. Using $a_1 = \alpha$ and $a_5 = \sqrt{1-\alpha^2}$,
we find $f(a_x) = \alpha^2 (1-\alpha^2)$, which attains its maximum value $1/4$ for $\alpha = 1/\sqrt{2}$.
This shows that a qubit-qutrit state is maximally entangled with respect to negativity, if and only if it is
of the form of eq(\ref{eq:5}) for $\alpha = 1/\sqrt{2}$.

Let us consider a specific pure state of a given degree of entanglement:
\begin{eqnarray} \label{eq:6}
\ket{\Phi_{1}} = \alpha \, \ket{0, 0} + \beta \, \ket{1, 1},
\end{eqnarray}
The normalization of the state vectors demands that $ |\alpha|^{2} + |\beta|^{2} = 1 $. The negativity for these states is given by $ 2 \, \alpha \, \beta$. This value will remain invariant for all pure entangled states obtained by applying local unitaries to eq(\ref{eq:6}). Therefore, we can characterize the set of pure entangled states for a given degree of entanglement by eq(\ref{eq:5}). Some other examples of such states are 
\begin{eqnarray}\label{eq:7}
\ket{\Phi_{2}^{\pm}} = \alpha \, \ket{0, 1} \pm \beta \, \ket{1, 2},
\end{eqnarray}

\begin{eqnarray}\label{eq:8}
\ket{\Phi_{3}^{\pm}} = \alpha \, \ket{0, 2} \pm \beta \, \ket{1, 0},
\end{eqnarray}

It is clear from eq(\ref{eq:2}) that the partial transpose of pure states in $ 2 \times 3$ systems can have only one negative eigenvalue. However, for mixed entangled states, there can be two possible negative eigenvalues of the partially transposed matrix \cite{M07}. For the qubit-qutrit system, we expect that there can be at most two negative eigenvalues. For the two-qubit domain, there is only one possible negative eigenvalue of the partially transposed density matrix \cite{STV97} for all entangled states.

\section{Modeling of qubit-qutrit and the process of disentanglement}

In this section, we model our qubit as a two-level atom in trap 1. Our qutrit is a three-level atom in another trap 2, located at a large distance in space from trap 1. We will consider our three-level atom in the V configuration. In V configuration, the only allowed  transition are between excited states and the ground state. The atomic transition between two excited states is not allowed. The further properties and experimental setup of the V configuration can be found in \cite{FS04}. Both the traps are taken to be in their vacuum states. Similar analysis has been done for two qubits \cite{J06} and two qutrits (two three-level atoms) \cite{DJ06}. Our principal system of two atoms are initially entangled with each other, while both traps serve as the environment. As the atoms are sufficiently separated that they no longer interact, we can concentrate only on the principal system in the master equation. Due to the spontaneous emission of the excited states, the atoms can decay even in the absence of photons present in the traps. This causes the decoherence and the degradation of entanglement in the principal system. We are interested in studying the dynamics of this entanglement.

The master equation for the two separated atoms, describing the dissipative part, is given by

\begin{eqnarray}\label{eq:9}
\frac{d \rho}{dt} = \Upsilon^{AB} \rho ,
\end{eqnarray}
where 
\begin{eqnarray}\label{eq:10}
\Upsilon^{AB}\rho = \frac{1}{2} \gamma (2 \sigma_{01}^{A}\rho \sigma_{10}^{A} - \sigma_{11}^{A} \rho - \rho \sigma_{11}^{A}) +     \frac{1}{2} \gamma_{2} (2 \sigma_{02}^{B} \rho \sigma_{20}^{B} - \sigma_{22}^{B}\rho - \rho \sigma_{22}^{B}) \nonumber \\  + \frac{1}{2} \gamma_{1} (2 \sigma_{01}^{B} \rho \sigma_{10}^{B} - \sigma_{11}^{B} \rho - \rho \sigma_{11}^{B}).
\end{eqnarray}
Here 
\begin{displaymath}
\sigma_{kl}^{A} = \sigma_{kl} \otimes \mathbf{I_{3}}, \quad \sigma_{kl}^{B} = \mathbf{I_{2}} \otimes \sigma_{kl}.
\end{displaymath}

The atomic operator $\sigma_{kl} = |k\rangle\langle l |$ takes an atom from the state $\ket{l}$ to the state $\ket{k}$. $\mathbf{I_{n}}$ is the $n \times n$ identity matrix. In the case of a two-level atom, the levels are described as $\ket{0}$ being the ground state and $\ket{1}$ as the excited state. For the three-level atom, the two excited levels are denoted by $\ket{2}$ and $ \ket{1}$, while the ground state is denoted by $ \ket{0} $. $ \gamma $ is the decay constant for the two level atom A, while $ \gamma_{2} $ and $ \gamma_{1} $ are the atomic decay constants of level $\ket{2}$ to level $\ket{0}$ and level $\ket{1}$ to level $\ket{0}$ for the three-level atom B, respectively. We identify the atomic states $\ket{2}$, $\ket{1}$, and $\ket{0}$ of the three-level atom, with the canonical basis vectors defined by
\begin{displaymath}
\ket{0}_{B} = \left( \begin{array}{c}0 \\ 0 \\ 1 \end{array}\right), \qquad \ket{1}_{B} = \left( \begin{array}{c}0 \\ 1 \\ 0 \end{array}\right), \qquad \ket{2}_{B} = \left( \begin{array}{c}1 \\ 0 \\ 0 \end{array}\right).
\end{displaymath}

Similarly for the two level atom A, the canonical basis vectors are given by 
\begin{displaymath}
\ket{0}_{A} = \left( \begin{array}{c}0 \\ 1 \end{array}\right), \qquad \ket{1}_{A} = \left( \begin{array}{c}1 \\ 0 \end{array}\right).
\end{displaymath}

Let us consider a general density matrix with respect to the basis $\ket{1} \otimes \ket{2}$, $\ket{1} \otimes \ket{1}$, $\ket{1} \otimes \ket{0}$, $\ldots$, $\ket{0} \otimes \ket{0}$,

\begin{eqnarray} \label{eq:11}
\mathbf{\rho} = \left ( \begin{array}{cccc}
\rho_{11} & \rho_{12} & \ldots & \rho_{16} \\
\rho_{21} & \rho_{22} & \ldots & \rho_{26} \\
\vdots & \vdots & \ddots & \vdots \\
\rho_{61} & \rho_{62} & \ldots & \rho_{66}
\end{array} \right).
\end{eqnarray}
For this general density matrix in our system of interest, there are $36$ equations of motion derived from eq(\ref{eq:9}), $24$ of them being uncoupled and easily solved. The remaining are the coupled equations. We have not solved them for the general case as the calculations are elementary but tedious but have considered particular input states.

\subsection{Dynamics of entanglement for eq(\ref{eq:6})}

Consider the density matrix of eq(\ref{eq:6}) and its time evolution. At time $t = 0$, the only non-zero matrix elements are given by
\begin{displaymath}
\rho_{22} (0)= \beta^{2}, \quad \rho_{26}(0) = \rho_{62}(0) = \pm \, \alpha \, \beta \quad \rho_{66}(0) = \alpha^{2}.
\end{displaymath}
The matrix elements after the interaction time $t$ are given by
\begin{displaymath}
\rho_{22} (t) = \beta^{2} \, e^{- (\gamma + \gamma_{1})t},
\end{displaymath}

\begin{displaymath}
\rho_{26} (t) = \rho_{62}(t) = \pm \, \alpha \, \beta \, e^{\frac{-(\gamma + \gamma_{1})t}{2}}, 
\end{displaymath}

\begin{displaymath}
\rho_{33}(t) = \beta^{2} \, (e^{- \gamma t}\, - \, e^{-(\gamma + \gamma_{1})t}),
\end{displaymath}

\begin{displaymath}
\rho_{55}(t) = \beta^{2} \, (e^{- \gamma_{1} t}\, - \, e^{-(\gamma + \gamma_{1})t}),
\end{displaymath}

\begin{displaymath}
\rho_{66} (t) = 1 - \, \beta^{2} \, (e^{- \gamma t} \, + \, e^{- \gamma_{1} t} \, - \, e^{- (\gamma + \gamma_{1})t}), 
\end{displaymath}
all the remaining matrix elements being zero. Taking the partial transpose of the matrix, and computing its eigenvalues, we have for the negativity,

\begin{eqnarray} \label{eq:12}
N_{1}(\beta) = \, \max \, \Biggl\{ \, 0, \, e^{-\, (\gamma \, + \, \gamma_{1})\,t} \, \biggl(\, \beta^{2} \, (\, 2 - e^{\gamma \,t} - e^{\gamma_{1}\,t}\,) \nonumber \\ + \quad \sqrt{\beta^{4}\, ( \,e^{2 \, \gamma \,t} + e^{2 \, \gamma_{1}\, t}\,) + (\,4 \, \beta^{2} - 6 \, \beta^{4}\,) \, e^{(\gamma \, + \, \gamma_{1})\,t}}\, \biggr) \, \Biggr\}.
\end{eqnarray}
Note immediately that at $ t = 0 $, the negativity is $ 2 \, \alpha \, \beta $. Let us take the maximally entangled state, that is,  $\alpha = \beta = 1/\sqrt{2}$. For this case, eq(\ref{eq:12}) reduces to
\begin{eqnarray}\label{eq:13}
N_{1} = \, e^{- (\, \gamma \, + \, \gamma_{1})\, t}. 
\end{eqnarray}
It is evident that maximally entangled states lose their entanglement asymptotically. Also for the parameter range of $0 < \beta \leq \frac{1}{\sqrt{2}}$, the states lose their entanglement in infinite time. But for $\frac{1}{\sqrt{2}} < \beta < 1$, the phenomenon of sudden death appears. However, this example is equivalent to pure states of two qubits, which exhibit sudden death \cite{J06}. Figure 1 show the contour plots for negativity (eq(\ref{eq:12})) for different values of $\beta$ in the range of sudden death.

\subsection{Dynamics of entanglement for eq(\ref{eq:7})}

Let us consider the density matrix of eq(\ref{eq:7}). This state is locally equivalent to state (\ref{eq:6}), and therefore has the same degree of entanglement. At $ t = 0 $, the non-zero elements are
\begin{displaymath}
\rho_{11}(0) = \beta^{2}, \quad \rho_{15}(0) = \rho_{51}(0) = \pm \, \alpha \, \beta \quad \rho_{55}(0) = \alpha^{2}.
\end{displaymath}
After the interaction with the cavities, the elements of the matrix are given by
\begin{displaymath}
\rho_{11} (t) = \beta^{2} \, e^{-(\, \gamma \, + \, \gamma_{2})\,t},
\end{displaymath}

\begin{displaymath}
\rho_{15} (t) = \rho_{51}(t) = \pm \, \alpha \, \beta \, e^{\frac{-(\,\gamma \, + \, \gamma_{1} \, + \, \gamma_{2})\,t}{2}},
\end{displaymath}

\begin{displaymath}
\rho_{33} (t) = \beta^{2} \, (\, e^{- \, \gamma \,t} \, - \, e^{-\, ( \, \gamma \, + \, \gamma_{2}) \, t}),
\end{displaymath}

\begin{displaymath}
\rho_{44} (t) = \beta^{2} \, (e^{- \, \gamma_{2} \, t} \, - \, e^{-( \, \gamma \, + \, \gamma_{2}) \, t}),
\end{displaymath}

\begin{displaymath}
\rho_{66} (t) = 1 - \alpha^{2} \, ( \, e^{- \gamma_{1} \, t} - \, \beta^{2} \, ( \, e^{- \, \gamma \, t} \, + \, e^{- \, \gamma_{2} \, t} - \, e^{-( \, \gamma \, + \, \gamma_{2}) \, t});
\end{displaymath}
all the remaining matrix elements are zero. Note the appearance of decay factors $\gamma_{1}$ and $\gamma_{2}$. This simply reflects the fact that in our initial pure state(\ref{eq:7}) both upper levels of the three-level atom are now involved. This can cause the quantum interference between levels $\ket{1}$ and $\ket{2}$ of our qutrit. So, before calculating negativity, a brief description of this interference is appropriate.
 
It is well known that the quantum interference occurs in three-level atoms in the V configuration \cite{FS04}. A measure of this interference is given by 

\begin{eqnarray}\label{eq:14}
k = \frac{\gamma_{1}}{\gamma_{2}}.
\end{eqnarray}
To see the effects of this interference, we demand that $\gamma_{1}\ll \gamma_{2} $. The interference is maximum for $\gamma_{1} = 0$. This interference has a profound effect on the process of disentanglement in the system of two entangled qutrits \cite{DJ06}. In the qubit-qutrit systems, it also affects the process of disentanglement in a similar manner. For $k = 1$, there is no quantum interference. As $ k $ decreases, the quantum interference increases and it is maximum for $ k = 0 $.

The negativity of the time evolved state (\ref{eq:7}) is given by

\begin{eqnarray} \label{eq:15}
N_{2} = e^{- \gamma_{2} \, t} \left( \beta^{2} ( e^{- \gamma \, t} -1 ) + \sqrt{\beta^{4} (-1 + e^{- \gamma \, t})^2 + 4 \alpha^{2} \beta^{2} e^{-(\gamma + \gamma_{1} - \gamma_{2})t} } \, \right).
\end{eqnarray}
Observe from this relation that for maximally entangled states, this negativity has its maximum value of 1 at $t = 0$, and the states becomes completely separable only at $ t = \infty$. But, the process of disentanglement is different from eq(\ref{eq:12}). Sudden death never occurs for any value of $ \beta $. However, the quantum interference may be used to control the process of disentanglement in such states. Figure 2 shows the behavior of eq(\ref{eq:15}) for zero  and maximum interference. The other locally equivalent pure state $\ket{\Phi_{2}^{'}} = \alpha \ket{02} + \beta \ket{11}$ exhibits the same dynamics.

We have shown that certain pure entangled states for a given degree of entanglement exhibit the sudden death of entanglement, while other locally equivalent pure states do not. This situation is similar to the two-qubit case \cite{J06}. However, the quantum interference is an additional nice feature of higher dimensions of Hilbert space. 

\section{Mixed states}

In this section, we consider an important class of mixed states for $2\times n$ systems \cite{CL03}. There it has been shown that an arbitrary state $\rho$ in a $2 \times 3$ quantum system can be transformed to a state of the form in equation(\ref{eq:16}) by local operations. This is a two-parameter class of states. For our system of interest, the states are given by

\begin{eqnarray}\label{eq:16}
\rho_{a, c} = c \, |\Psi^{-}\rangle\langle\Psi^{-}| \, + \, b \, (\, |\Psi^{+}\rangle\langle\Psi^{+}| \, + \, |00\rangle\langle00| \, + \,  |11\rangle\langle11| \, ) \, + \, a \nonumber \\ (\, |02\rangle\langle02| \, + \, |12\rangle\langle12| \, ),
\end{eqnarray}
where $\ket{\Psi^{\pm}} = \frac{1}{\sqrt{2}} \ket{0,1} \pm \ket{1,0}$, and the unit trace constrains the parameters to satisfy the relation

\begin{displaymath}
2 a + 3 b + c = 1. 
\end{displaymath}
The non-zero matrix elements, at $t = 0$, are given by 
\begin{displaymath}
\rho_{11}(0) = a, \quad \rho_{22}(0) = b, \quad \rho_{33}(0) = \frac{b + c}{2},
\end{displaymath}

\begin{displaymath}
\rho_{35} = \rho_{53}(0) = \frac{b - c}{2}, \quad \rho_{44}(0) = a, \quad \rho_{55}(0) = \frac{b + c}{2}, \quad \rho_{66}(0) = b.
\end{displaymath}
The non-zero elements of the time-dependent matrix are given by
\begin{displaymath}
\rho_{11}(t) = a \, e^{- \, (\gamma \, + \, \gamma_{2})\,t},
\end{displaymath}

\begin{displaymath}
\rho_{22}(t) = b \, e^{-\, (\gamma \, + \, \gamma_{1})\, t},
\end{displaymath}

\begin{displaymath}
\rho_{33}(t) = \frac{e^{-\gamma \, t}}{2} \Bigl(1\, - 2 \, (b \, e^{-\gamma_{1}\, t} \, + \, a \, e^{-\gamma_{2}\, t}\, )\, \Bigr), 
\end{displaymath}

\begin{displaymath}
\rho_{35}(t) = \rho_{53}(t) = \left(\frac{b\, -\, c}{2}\right) \, e^{\frac{-\, (\gamma \, + \, \gamma_{1})\, t}{2}},
\end{displaymath}

\begin{displaymath}
\rho_{44}(t) = a \, (2 \, e^{-\gamma_{2}\, t} \, - \, e^{-\, (\gamma \, + \, \gamma_{2})\, t}\, ),
\end{displaymath}

\begin{displaymath}
\rho_{55}(t) = \left(\frac{3\, b\, + \, c}{2}\right) e^{- \gamma_{1}\, t} \, - \, b \, e^{-\, (\gamma \, + \, \gamma_{1})\, t},
\end{displaymath}

\begin{displaymath}
\rho_{66}(t) = 1 + \frac{e^{-(\gamma + \gamma_{1} + \gamma_{2})t}}{2} \,\left( e^{\gamma_{1}t}(2 a - 4 a e^{\gamma t}) \, + \, e^{\gamma_{2}t} ( 2 b + (-3 b - c)e^{\gamma t} -  e^{\gamma_{1}t}) \right). 
\end{displaymath}
The expression of negativity in this case is lengthy but easily obtainable by linear algebra packages. Rather than reproducing it here, we present the main results as plots.

Let us fix the parameter $b$ in the above matrix elements and study the sudden death in this class of mixed states. For $b = 0.02$, the sudden death occurs in the range $0 < c \lesssim 0.302$ provided that interference is zero, that is, $\gamma_{1} \approx \gamma_{2}$. However, the sudden death is delayed when interference increases and for maximum interference, the sudden death occurs in the range $0 < c \lesssim 0.2775$. Figures 3 show contour plots of negativity versus dissipation factors for different values of the parameter $ c $.

Similarly for the fixed parameter $b = 0.06$ and zero interference, the sudden death occurs in the range $ 0 < c \lesssim 0.5493$. However, the interference delays the sudden death and for maximum interference the sudden death occurs in range $0 < c \lesssim 0.46295$. Figures 4 show contour plots of negativity for different values of the parameter $c$. We have also observed the sudden death in this class of states when the parameter $a$ ($b$) is zero for certain range of parameter $c$. We expect the same phenomena for the parameter $c = 0$.

In the particular set-up we have analyzed, all entangled states, whether pure or mixed, end at $ t = \infty $ in the particular pure state given by 

\begin{eqnarray}\label{eq:17}
\rho_{\infty} = \left( \begin{array}{cccccc} 0 & 0 & 0  & 0 & 0 & 0 \\
                                             0 & 0 & 0 & 0 & 0 & 0  \\
                                             0 & 0 & 0 & 0 & 0 & 0 \\
                                             0 & 0 & 0 & 0 & 0 & 0 \\  
                                             0 & 0 & 0 & 0 & 0 & 0 \\
                                             0 & 0 & 0 & 0 & 0 & 1 \\ \end{array} \right).
\end{eqnarray}

\section{Conclusions}

We have investigated the phenomenon of entanglement sudden death for qubit-qutrit entangled states. We have considered both pure and mixed states. We have shown that ESD can happen both for pure and mixed entangled states for a qubit-qutrit system. Certain pure entangled states exhibit the sudden death for a particular range of a single parameter. Other pure states, obtained by applying local unitaries to this particular state, loose their entanglement at infinity. This observation is similar to two-qubit pure states. However, in $2\times 3$ and other higher dimensions of Hilbert space, the quantum interference controls the dynamics of entanglement. We have also observed this effect of sudden death in a special class of mixed states. This effect of ESD has been reported for two qubits and two qutrits previously. Hence, it seems that ESD is common in all dimensions of Hilbert spaces. During the completion of our work, we have found that similar conclusions have been arrived at but working in a different way \cite{AJ07}.

Having sufficient evidence for this striking phenomenon, the future challenge is to work out effective techniques to tackle ESD \cite{RMG07}. Loss of entanglement seriously endangers experimental implementation of quantum information processors, so that prolonging it will be of considerable interest.

\ack

The authors thank Gernot Alber for many helpful suggestions and discussions. ARPR thanks the Theoretische Quantenphysik group at the Technische Universit\"{a}t Darmstadt, for its hospitality during the course of this work. M. Ali acknowledges financial support by the Higher Education Commission, Pakistan, and the Deutscher Akademischer Austauschdienst, Bonn. K. Ranade is supported by a graduate-student scholarship (Promotionsstipendium) of the Technische Universit\"{a}t Darmstadt.


\begin{figure}
\begin{center}
\includegraphics[width = 6cm]{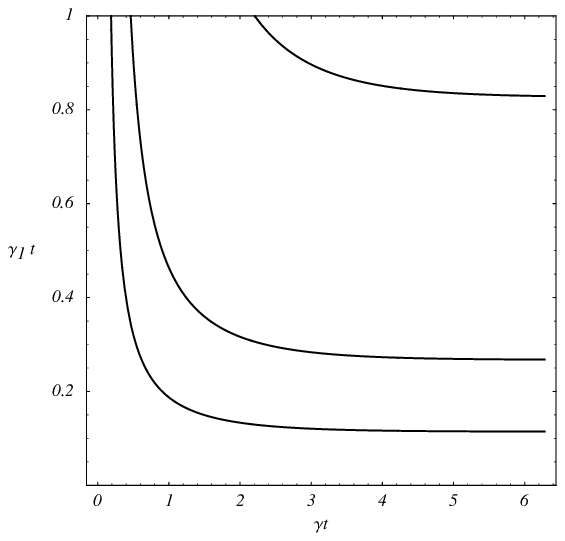} 
\end{center}
\caption{Time dependence of negativity eq(\ref{eq:12}), for different values of $\beta \in [ \frac{1}{\sqrt{2}}, 1 ]$. The lowest, middle and upper curves are for $\beta = 0.95$, $0.9$ and $0.8$ respectively. The each curve is a boundary between entangled and separable states. The states above and on the boundary are separable.}
\end{figure}

\begin{figure}
\begin{center}
\includegraphics[width = 14cm]{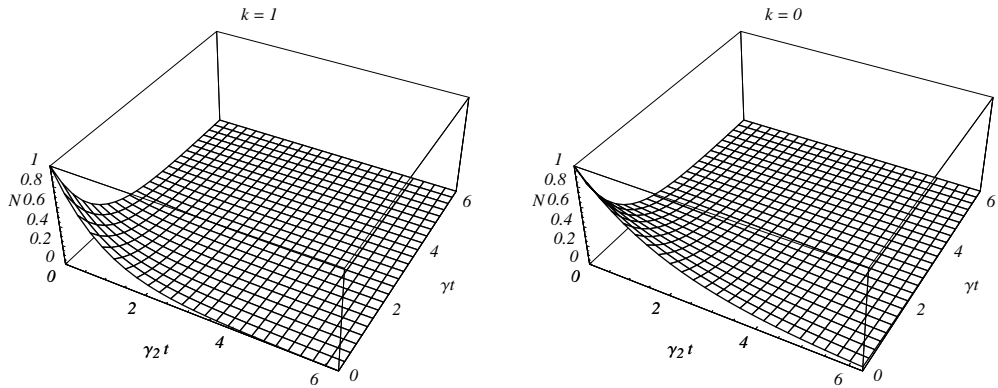} 
\end{center}
\caption{Time dependence of negativity for a maximally entangled state for zero and maximum interference, that is, $ k = 1 $ and $0$ respectively.}
\end{figure}

\begin{figure}
\begin{center}
\includegraphics[width = 6cm]{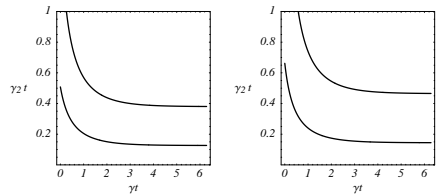} 
\end{center}
\caption{Contour plots of negativity is plotted for the fixed parameter $b = 0.02$. The lower and upper curves correspond to $c = 0.15$ and $0.2$ respectively. The left graph is for $ k = 1$ and right for $k = 0$. The region above and on each curve is that of sudden death.}
\end{figure}

\begin{figure}
\begin{center}
\includegraphics[width = 6cm]{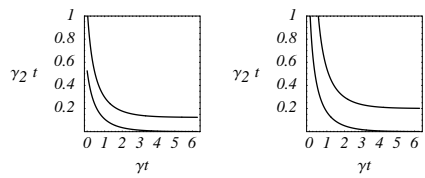} 
\end{center}
\caption{Same as Fig. 3 for $ b = 0.06 $. The lower and upper curves are for $c = 0.25$ and $0.4$, respectively.}
\end{figure}

\end{document}